\documentclass[final,5p,times,twocolumn,authoryear]{elsarticle}

\usepackage{amsmath}
\usepackage{amssymb}
\usepackage[colorlinks,citecolor=blue,linkcolor=blue,urlcolor=blue]{hyperref}

\newcommand{\degr}{\ensuremath{^\circ}}
\newcommand{\farcs}{\mbox{$.\!\!^{\prime\prime}$}}
\newcommand{\tablefootmark}[1]{\ensuremath{^{\mathrm{#1}}}}
\newcommand{\tablefoottext}[2]{\ensuremath{^{\mathrm{#1}}}\,#2}
\newcommand{\tablefoot}[1]{\par\vspace{3pt}\begin{minipage}{\linewidth}\footnotesize\raggedright\textit{Notes.} #1\end{minipage}}

\journal{Journal of High Energy Astrophysics}

\begin{document}

\begin{frontmatter}

\title{Kinematic and gamma-ray limits on a binary origin for the offset broad H$\alpha$ line of PKS\,0903$-$57}

\author[ucm]{Alberto Dom\'inguez\corref{cor1}}
\ead{alberto.d@ucm.es}
\cortext[cor1]{Corresponding author}
\author[ucm]{Adithiya Dinesh}
\author[clemson]{Pablo Pe\~nil}
\author[clemson]{Alba Rico}
\author[wurzburg,desy]{Sara Buson}
\author[clemson]{Marco Ajello}

\affiliation[ucm]{organization={IPARCOS and Department of EMFTEL, Universidad Complutense de Madrid},
            city={Madrid},
            postcode={E-28040},
            country={Spain}}
\affiliation[clemson]{organization={Department of Physics and Astronomy, Clemson University, Kinard Lab of Physics},
            city={Clemson},
            postcode={29634-0978},
            state={SC},
            country={USA}}
\affiliation[wurzburg]{organization={Julius-Maximilians-Universit\"at W\"urzburg, Fakult\"at f\"ur Physik und Astronomie},
            addressline={Emil-Fischer-Str. 31},
            city={W\"urzburg},
            postcode={D-97074},
            country={Germany}}
\affiliation[desy]{organization={Deutsches Elektronen-Synchrotron DESY},
            addressline={Platanenallee 6},
            city={Zeuthen},
            postcode={D-15738},
            country={Germany}}

\begin{abstract}
PKS\,0903$-$57 is a flat-spectrum radio quasar at $z=0.2621$ whose broad H$\alpha$ line is
offset by $1460\pm36$\,km\,s$^{-1}$ from the systemic redshift, and a supermassive binary black
hole (SMBBH) is one of the interpretations proposed for this offset. We test the physical
viability of that interpretation through its orbital kinematics and through an 18\,yr
\textit{Fermi}-LAT gamma-ray periodicity search. We evaluate circular-binary solutions consistent
with the offset for a black hole mass derived from the broad H$\alpha$ luminosity and width,
requiring the broad-line region to remain bound to the moving hole. We analyze \textit{Fermi}-LAT
Light Curve Repository data with five periodogram statistics: a generalized Lomb-Scargle
periodogram, and a Lomb-Scargle periodogram and three morphology-aware templates on the singular
spectrum analysis (SSA) oscillatory component. For the fiducial mass and broad-line region
radius, the offset and the confinement of the broad-line region require an observer-frame
orbital period of at least 230\,yr, and at least 60\,yr allowing for their systematic
uncertainties; a 2.1\,yr period would need a broad-line region more than 1\,dex smaller than
expected. The raw-flux periodogram is consistent with red noise. SSA isolates a 2.12\,yr
oscillatory component (global $p=0.15$, $1.5\sigma$, for a stationary red-noise null); a joint
test across the five statistics gives $p=0.27$ ($1.1\sigma$), or $p=0.063$ ($1.9\sigma$) for a
null following the observed variability envelope. The templates show no morphological
preference. The candidate is confined to the second half of the baseline and depends on
individual flares, and wavelet ridge tracking finds a coherent modulation whose period wanders
between about 1.5 and 2.3\,yr. Neither the gamma-ray light curve nor the orbital dynamics
provides positive evidence for a binary origin of the line offset. In the adopted model, no
binary hosting the broad line has an orbital period comparable to the gamma-ray candidate. A
wide binary remains possible and would not produce a detectable orbital modulation on the LAT
timescale.
\end{abstract}


\end{frontmatter}

\section{Introduction}
\label{sec:intro}

Velocity offsets between the broad and narrow emission lines of active galactic nuclei (AGN)
are among the few observational signatures that can point directly at a supermassive binary
black hole (SMBBH). In a binary, the broad-line region (BLR) may remain bound to one hole
while the narrow-line region traces the systemic velocity of the host, producing a
persistent offset \citep[e.g.][]{dotti2009,kelley2021}. The same signature can, however, be
produced by a black hole recoiling after coalescence \citep[e.g.][]{campanelli2007,
chiaberge2018} or by a BLR made of infalling clumps shed from the dusty torus
\citep[e.g.][]{wang2017}.

\citet{goldoni2024} reported such an offset in PKS\,0903$-$57. Using VLT/FORS2 spectroscopy
that resolves the blazar from a foreground Galactic star at $0\farcs67$, they measured a firm
redshift $z=0.2621\pm0.0006$ from five narrow lines, classified the source as a flat-spectrum
radio quasar (FSRQ) from a broad H$\alpha$ line of equivalent width $7.94\pm0.17$\,\AA{} and
full width at half maximum $4020\pm30$\,km\,s$^{-1}$, and found the broad H$\alpha$ centroid
redshifted by $1460\pm36$\,km\,s$^{-1}$ relative to the narrow lines, at approximately
$40\sigma$. They raised all three interpretations above and noted that, for a sub-parsec
binary, no velocity variation would be detectable over five years \citep{kelley2021}, and the offset alone therefore does not discriminate among them.

The three scenarios do make different predictions for the temporal behavior of the jet. A
recoiling hole carries its disk and jet with it and produces no periodic signal. An infalling clumpy
BLR produces irregular line-profile variability but no jet modulation. A binary can imprint a
period on the jet through modulated accretion, whose observed period is the orbital period
redshifted, or through geometric reorientation of a relativistically moving emitting region,
for which the mapping between orbital and observed timescales is model dependent. A gamma-ray
periodicity search is therefore a test of the binary hypothesis, and PKS\,0903$-$57 has been
monitored by the \textit{Fermi} Large Area Telescope (LAT) for 18\,yr.

That test has to be made carefully. Red noise alone generates apparent periodicities at one
and a half to two and a half cycles per baseline \citep[e.g.][]{vaughan2016,covino2018},
strong flares distort or manufacture periodic structure \citep[e.g.][]{penil2025flares}, and
data gaps introduce spurious features near one year \citep[e.g.][]{penil2025gaps}. The most
recent systematic reanalyses find that almost no previously reported gamma-ray periodicity
survives a proper trials correction \citep[e.g.][]{penil2026,rico2025}.

This paper is organized as a two-stage test. Section~\ref{sec:kin} evaluates whether the
measured line offset admits a binary solution at all, independently of any gamma-ray
information. Section~\ref{sec:data} describes the gamma-ray data and Sect.~\ref{sec:analysis}
the time-domain analysis, carried out with complementary estimators.
Section~\ref{sec:robust} reports the robustness and systematics tests, Sect.~\ref{sec:disc}
discusses the alternatives, and Sect.~\ref{sec:concl} summarizes. We use a flat cosmology
with $H_0=70$\,km\,s$^{-1}$\,Mpc$^{-1}$ and $\Omega_{\rm M}=0.3$. Periods are quoted in the
observer frame unless stated otherwise.

\section{Kinematic constraints on a binary}
\label{sec:kin}

\subsection{Black hole mass}
\label{sec:mass}

The optical continuum of PKS\,0903$-$57 is jet dominated, as \citet{goldoni2024} demonstrate, and the continuum luminosity cannot be used as a mass proxy. We therefore use the broad
H$\alpha$ relation of \citet{greene2005} with the luminosity and width measured by
\citet{goldoni2024}, $L_{{\rm H}\alpha}=1.0(\pm0.5)\times10^{43}$\,erg\,s$^{-1}$ and
${\rm FWHM}=4020\pm30$\,km\,s$^{-1}$, which gives
\begin{equation}
M_{\rm BH}=1.25\,(0.85\text{--}1.56)\times10^{8}\,M_{\odot},
\end{equation}
where the range propagates the quoted luminosity uncertainty only. A single-epoch
virial estimate also carries roughly 0.4 dex of systematic scatter from the virial
factor and broad-line region geometry \citep[e.g.][]{shen2013}, which we do not include in the
quoted range; the sensitivity of the joint constraint of Sect.~\ref{sec:blr} to the measured
H$\alpha$ mass is evaluated there.

\subsection{Orbital solutions}

Consider a binary in which the broad-line region is bound to a hole of mass $m$ with
companion $M$ and total mass $M_{\rm tot}$. We define the inclination $i$ as the angle
between the orbital angular momentum and the line of sight; $i=0\degr$ is then a face-on orbit and the line-of-sight velocity of the line-emitting hole is
$v_1\sin i\cos\phi$ with $\phi$ the orbital phase. For a circular orbit
$v_1=(M/M_{\rm tot})\,v_{\rm rel}$ and
\begin{equation}
P_{\rm orb}=\frac{2\pi G M_{\rm tot}}{v_{\rm rel}^{3}},
\qquad
a=\frac{G M_{\rm tot}}{v_{\rm rel}^{2}} .
\label{eq:porb}
\end{equation}

\begin{figure}
\centering
\includegraphics[width=\linewidth]{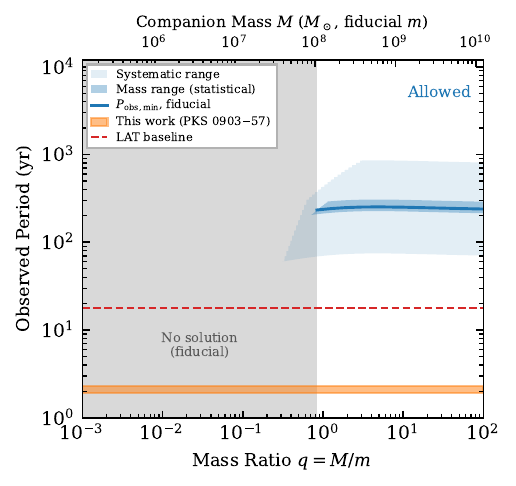}
\caption{Shortest observer-frame period compatible with both conditions of
Eq.~(\ref{eq:joint}), as a function of mass ratio, for a circular orbit. The curve is for the
fiducial $m=1.25\times10^{8}\,M_{\odot}$ and $R_{\rm BLR}=38$\,mpc; the dark band spans the
statistical mass range of Sect.~\ref{sec:mass}, and the light band the systematic range of
$\pm0.4$\,dex in mass and $\pm0.22$\,dex in radius (Table~\ref{tab:sys}). The upper axis gives
the companion mass for the fiducial $m$. Periods above the curve are allowed; no fiducial
configuration exists in the shaded region at left. The gamma-ray candidate and the length of
the LAT baseline are shown for comparison.}
\label{fig:binary}
\end{figure}

Because the spectroscopy is a single epoch at an unknown phase, $|\cos\phi|\le1$ gives $v_1\sin i\ge\Delta v$, and the velocity inferred by identifying the measured offset with
the full projected amplitude is a lower bound on $v_{\rm rel}$. Since
$P_{\rm orb}\propto v_{\rm rel}^{-3}$ and $a\propto v_{\rm rel}^{-2}$, the periods and
separations obtained this way are therefore upper bounds, and the inspiral times, which
scale as $a^{4}$, are upper bounds as well. Table~\ref{tab:binary} and
Fig.~\ref{fig:binary} evaluate Eq.~(\ref{eq:porb}) on that basis for equal masses of
$1.25\times10^{8}\,M_{\odot}$ each.

In models where the binary orbital angular momentum is assumed to be aligned with the jet, the
orbit is seen nearly face-on and $i\lesssim5\degr$ for a blazar. The implied observed period is then at most 0.22\,yr and
the separation at most 0.96\,mpc. If instead the orbit is edge-on, the observed period is
at most 335\,yr, almost twenty times the LAT baseline. For an equal-mass circular binary an observed period of 2.12\,yr
corresponds to $i\simeq11\degr$, with $a\le4.3$\,mpc.

\begin{table}
\caption{Equal-mass circular binary solutions for
$M_{\rm tot}=2.5\times10^{8}\,M_{\odot}$ and $\Delta v=1460$\,km\,s$^{-1}$.}
\label{tab:binary}
\centering
\footnotesize
\begin{tabular}{@{}lcccc@{}}
\hline\hline
$i$ & $v_{\rm rel}$ & $P_{\rm obs}$ & $a$ & $t_{\rm GW}$ \\
(deg) & (km\,s$^{-1}$) & (yr) & (mpc) & (yr) \\
\hline
5    & $3.4\times10^{4}$ & $\le0.22$ & $\le0.96$ & $\le1.2\times10^{2}$ \\
10   & $1.7\times10^{4}$ & $\le1.75$ & $\le3.8$  & $\le3.1\times10^{4}$ \\
11   & $1.5\times10^{4}$ & $\le2.33$ & $\le4.6$  & $\le6.6\times10^{4}$ \\
20   & $8.5\times10^{3}$ & $\le13.4$ & $\le14.8$ & $\le7.1\times10^{6}$ \\
63\tablefootmark{a} & $3.3\times10^{3}$ & $\le236$ & $\le100$ & $\le1.5\times10^{10}$ \\
90   & $2.9\times10^{3}$ & $\le335$  & $\le126$  & $\le3.8\times10^{10}$ \\
\hline
\end{tabular}
\tablefoot{All entries are upper bounds, because the single-epoch offset is a lower
bound on the projected velocity amplitude (Sect.~\ref{sec:kin}). $P_{\rm obs}=(1+z)
P_{\rm orb}$ for an orbital (accretion-driven) modulation.
\tablefoottext{a}{Smallest inclination for which the broad-line region fits inside the
Roche lobe of its host in this equal-mass case; the general result is derived in
Sect.~\ref{sec:blr}.}}
\end{table}

\subsection{Mass ratio and eccentricity}
\label{sec:qe}

That inclination is not a unique solution, and we do not claim it as one. Relaxing the
mass ratio changes the answer substantially. Writing $q=M/m$ for the companion of the
line-emitting hole, Eq.~(\ref{eq:porb}) gives
$P_{\rm orb}=2\pi G m\,q^{3}\sin^{3}i/[\Delta v^{3}(1+q)^{2}]$: a heavier companion raises $v_1/v_{\rm rel}$ and lengthens the period at fixed inclination. At $i=5\degr$, for example, a system with $q=3.8$ reproduces an observed
period of 2.12\,yr with $a\le5.8$\,mpc. Year-scale periods are therefore available in the unconstrained circular
kinematic problem once the mass ratio is free; Sect.~\ref{sec:blr} shows that the
broad-line-region confinement requirement removes these compact solutions.

Eccentricity lengthens the implied periods further. If the measured offset is a pericenter
speed, $v_{\rm p}^{2}=GM_{\rm tot}(1+e)/[a(1-e)]$ allows a semimajor axis larger by
$(1+e)/(1-e)$ and a period larger by $[(1+e)/(1-e)]^{3/2}$, a factor of 8 at $e=0.6$ and 83 at
$e=0.9$; this is an upper bound on the effect, because the speed may be sampled away from
pericenter. The kinematics alone therefore do not isolate a narrow region of parameter space,
but they tie the observed period to the orbital separation, which the next section constrains
physically.

\subsection{Can the broad-line region fit inside the orbit?}
\label{sec:blr}

The offset of the broad H$\alpha$ centroid is an observational result; that it is the orbital
velocity of one component of a binary is the hypothesis under test. A displaced centroid can
also arise in a single-hole broad-line region with non-axisymmetric emission, which multi-line
profiles can help to discriminate \citep[e.g.][]{runnoe2025}. If the hypothesis holds, the
broad-line region must be hosted by the moving hole and fit inside its Roche lobe, and it is
this consequence that we test. Imposing it together with the velocity constraint turns the
loose kinematics of Sect.~\ref{sec:qe} into a limit.

Inverting the \citet{greene2005} relation for the measured
$L_{{\rm H}\alpha}=1.0\times10^{43}$\,erg\,s$^{-1}$ gives a thermal continuum
$L_{5100}=1.7\times10^{44}$\,erg\,s$^{-1}$, in agreement with the value quoted by
\citet{goldoni2024}. The radius-luminosity relation of \citet{bentz2013}, which is calibrated
on H$\beta$ reverberation mapping and which we adopt as a characteristic broad-line region
radius for the H$\alpha$-emitting gas, then gives
\begin{equation}
R_{\rm BLR}\simeq45\ {\rm light\ days}\simeq38\ {\rm mpc},
\end{equation}
with roughly 0.2 dex of intrinsic scatter and the additional uncertainty of an indirectly
inferred continuum. This is an emissivity-weighted radius and not an outer edge. Requiring it to be contained is therefore a conservative necessary condition, not an exact criterion: the test shows that the characteristic broad-line region inferred from the observed
line cannot be associated with one component under the assumed model, not that no H$\alpha$
photons could be produced in such a binary.

We identify the hole of mass $m$ with the one whose mass the H$\alpha$ virial estimate
constrains, and require simultaneously that
\begin{equation}
f_{\rm E}(m/M)\,a\,(1-e)\ \ge\ R_{\rm BLR},
\qquad
\max_{\phi}\,v_1\sin i\ \ge\ \Delta v,
\label{eq:joint}
\end{equation}
where $f_{\rm E}$ is the Eggleton Roche-lobe factor \citep{eggleton1983}, with the standard
approximation of evaluating the lobe at the pericenter separation, and the second condition
asks only that the orbit be able to produce the observed offset at some phase and inclination.
The Eggleton expression is derived for circular, synchronous binaries, and the limitations of
the pericenter substitution have been examined by \citet{sepinsky2007}; they are modest and do
not affect the qualitative conclusion. The first condition sets a minimum
separation and therefore a minimum period; the second excludes mass ratios for which the
orbital speed at that separation is too small.

We search companion masses $10^{-3}\le q\le10^{2}$, that is
$1.25\times10^{5}\le M/M_{\odot}\le1.25\times10^{10}$ for the nominal $m$, the upper end being
comparable to the most massive black holes known. For the nominal
$m=1.25\times10^{8}\,M_{\odot}$ and a circular orbit, no configuration with $q<0.84$ satisfies
both conditions, and the shortest observed-frame period allowed by those that do is 233\,yr, at
$q=0.84$ and $a=96$\,mpc (Fig.~\ref{fig:binary}), corresponding to a source-frame orbital
period of 185\,yr. Above $q=0.84$ the minimum varies by less than 10 per cent, because a
heavier companion raises the total mass and shrinks the Roche lobe of the line-emitting hole
in compensation; extending the domain to the unphysical $q=10^{4}$ lowers it only to an
asymptotic 230\,yr. Across the statistical mass range of Sect.~\ref{sec:mass} the minimum
observed period is 205 to 290\,yr (163 to 230\,yr in the source frame). The equal-mass case,
$f_{\rm E}=0.38$, gives $a\ge100$\,mpc and $P_{\rm obs}\ge237$\,yr.

These numbers are fiducial. They assume the H$\alpha$ virial mass and the characteristic
radius above, and neither the 0.4\,dex systematic uncertainty of single-epoch masses nor the
scatter of the radius-luminosity relation is included in them. We therefore repeat the
calculation over $\pm0.4$\,dex in $m$ and $\pm0.22$\,dex in $R_{\rm BLR}$, the latter combining
the 0.19\,dex intrinsic scatter of the \citet{bentz2013} relation with about 0.1\,dex from the
indirect conversion of $L_{{\rm H}\alpha}$ into $L_{5100}$ (Table~\ref{tab:sys} and the light
band in Fig.~\ref{fig:binary}). The minimum observed period scales approximately as
$R_{\rm BLR}^{3/2}m^{-1/2}$ and spans 60 to 810\,yr over that range, the shortest value
corresponding to a mass 0.4\,dex above and a radius 0.22\,dex below the fiducial ones. The
period floor is therefore uncertain by about an order of magnitude, but it stays above the LAT
baseline at every corner of the range. The comparison that matters for the gamma-ray candidate
is more robust: an observed period of 2.12\,yr would require a characteristic broad-line region
of 2.0\,mpc for the fiducial mass, or 2.8\,mpc for a mass 0.4\,dex higher, that is 1.1 to
1.3\,dex below the radius-luminosity prediction and five to six times its combined scatter.

\begin{table}
\caption{Shortest observer-frame period (yr) allowed by Eq.~(\ref{eq:joint}) for a circular
orbit, over the systematic ranges of black hole mass and broad-line region radius.}
\label{tab:sys}
\centering
\footnotesize
\begin{tabular}{@{}lccc@{}}
\hline\hline
 & $R_{\rm BLR}=23$\,mpc & 38\,mpc & 63\,mpc \\
\hline
$m=0.50\times10^{8}\,M_{\odot}$ & 177 & 379 & 811 \\
$m=1.25\times10^{8}\,M_{\odot}$ & 105 & 233 & 512 \\
$m=3.14\times10^{8}\,M_{\odot}$ & 60 & 136 & 304 \\
\hline
\end{tabular}
\tablefoot{Rows span $\pm0.4$\,dex in mass and columns $\pm0.22$\,dex in radius around the
fiducial values (central entry). Source-frame orbital periods are smaller by $1+z=1.262$. The
minimum is taken over $10^{-3}\le q\le10^{2}$.}
\end{table}

Because $v_1\sin i$ decreases with separation, the velocity condition in
Eq.~(\ref{eq:joint}) also caps the separation, and therefore the period, from above at fixed
$q$. For equal masses the allowed window is 237--335\,yr (Table~\ref{tab:binary}); the ceiling
grows roughly as $q^{3}$, already exceeding $4.6\times10^{3}$\,yr at $q=5$, and never excludes
periods long compared with the LAT baseline. The minima quoted above are unaffected.

Within the adopted prescription eccentricity only strengthens the limit. Because the Roche lobe
is set at pericenter, the minimum separation scales as $(1-e)^{-1}$, and repeating the
optimization over $q$ at fixed eccentricity gives minimum observed periods of 233, 390, 886
and $7.0\times10^{3}$\,yr for $e=0$, 0.3, 0.6 and 0.9. The circular case therefore defines the
shortest-period envelope, and the unknown orbital phase of a single-epoch measurement cannot be
used to shorten it.

The conclusion concerns an orbital modulation, that is one whose observed period is
$(1+z)P_{\rm orb}$, such as one driven by the orbital variation of the accretion rate. For such
a modulation, and within the model in which the broad-line region is bound to one component, no
binary capable of producing a gamma-ray modulation on a timescale comparable to the observed
2.12\,yr candidate can simultaneously reproduce the H$\alpha$ velocity offset and contain the
characteristic broad-line region; for the fiducial mass and radius the shortest allowed
observed period exceeds the 18\,yr baseline by more than an order of magnitude, and over the
systematic range of Table~\ref{tab:sys} by at least a factor of three. A geometric modulation
imposed on a relativistically moving emitting region need not obey that mapping, and we do not
use this argument against such scenarios; for those, the relevant evidence is the time-domain
analysis of Sect.~\ref{sec:analysis}.

\section{Gamma-ray data}
\label{sec:data}

We use the \textit{Fermi}-LAT Light Curve Repository \citep[LCR;][]{abdollahi2023}. The
source is 4FGL\,J0904.9$-$5734, the association adopted in the 4FGL catalog
\citep{abdollahi2020} and used in the flare reports of \citet{buson2020} and
\citet{lamura2022}; its 4FGL variability index is 8488. The 4FGL position agrees with the
VLBI radio position quoted by \citet{goldoni2024}.

The analysis combines a generalized Lomb-Scargle periodogram (GLS), a Lomb-Scargle periodogram
(LSP) of the oscillatory component isolated by singular spectrum analysis (SSA), the template
periodogram of \citet{maldonado2026}, and the weighted wavelet Z-transform (WWZ) ridge analysis
of \citet{dinesh2026}. For each we adopt the data conventions of the work that introduced it.

For the harmonic analysis we follow \citet{penil2026} and \citet{rico2025}: 30-day binned
light curves in 0.1--100\,GeV with the photon index fixed to its catalog value, bins with
${\rm TS}<4$ (a $2\sigma$ detection) treated as upper limits and removed, and sources
required to have fewer than 50 per cent missing bins. PKS\,0903$-$57 retains 208 of 221 bins,
a gap fraction of 5.9 per cent, over 18.1\,yr (MJD\,54698--61298). This is the light curve
analyzed by \citet{dinesh2026}, whose binning, photon-index and detection conventions are the
same, and it passes the sampling criteria of that work.

For the template analysis we follow \citet{maldonado2026}: 7-day binned light curves in
0.1--100\,GeV with a fixed photon index over MJD\,54686.5--61105.5, bins with ${\rm TS}<2$
treated as non-detections and removed, and sources required to have fewer than 30 per cent
non-detections. That work also introduces three data-quality cuts, which reject bins whose
likelihood fit did not converge, bins with ${\rm TS}\ge2$ whose flux uncertainty is zero or not
finite, and bins whose ratio $F/\sqrt{\rm TS}$ exceeds ten times its median over the light
curve, a signature of an unconstrained likelihood fit. PKS\,0903$-$57 retains 721 of 907 bins,
a gap fraction of 20.5 per cent, and therefore satisfies the selection criteria of that study.
It is not among the three sources of that study with a per-source significance above $3\sigma$,
nor among the 14 with a local significance of at least $3\sigma$, a number consistent with the
12.1 expected by chance in that sample.

We apply the same three cuts to every product used in this paper. None of them removes a bin
from the fixed-index 30-day or 7-day light curves of PKS\,0903$-$57, and the harmonic analysis
is therefore unaffected.

Removing the upper limits, without substituting them, is what all three studies do, and it is
the convention under which the significance calibrations of \citet{penil2025gaps} were
established. Both gap fractions lie well inside the regime in which \citet{penil2025gaps}
find period recovery to be reliable.

\begin{figure}
\centering
\includegraphics[width=\linewidth]{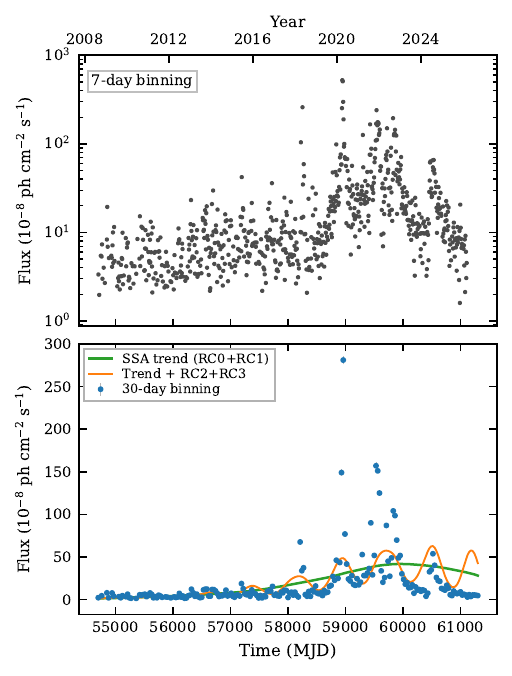}
\caption{\textit{Fermi}-LAT light curves of PKS\,0903$-$57 from the LCR. Upper panel:
7-day binning used for the blind-search template analysis, after the quality cuts of
\citet{maldonado2026}. Lower panel: 30-day binning used for the harmonic
analysis, with the SSA trend and the sum of trend and oscillatory component superimposed,
obtained with the algorithm of Sect.~\ref{sec:harm}. The variability amplitude changes
markedly between the quiet first decade and the active second half of the record.}
\label{fig:lc}
\end{figure}

The light curves are shown in Fig.~\ref{fig:lc}. The source is faint and weakly variable
during the first decade and becomes both brighter and more strongly variable after
approximately MJD\,58000. Four flaring episodes dominate the record, at
MJD\,$\approx$\,58250, 58950, 59530 and 59830, the second reaching 30 times the median 30-day
flux. That 2020 outburst led to the discovery of the source at very high energies and has been
analyzed in detail by the H.E.S.S. Collaboration \citep{hess2026}, who find strong variability and spectral evolution on
timescales far shorter than any considered here. They also find that the absence of
absorption beyond that expected from the extragalactic background light favors an emission
region outside the broad-line region, which we note as source-specific context and not as part of the argument of Sect.~\ref{sec:blr}. The variance is strongly concentrated in the second half of the baseline, which is the
single most important property of this data set for what follows.

\section{Gamma-ray time-domain analysis}
\label{sec:analysis}

\subsection{Null hypothesis and significance}
\label{sec:sig}

Analytic false-alarm probabilities assume white noise and do not apply. Every result below is
evaluated against Monte Carlo realizations to which the full analysis chain, gap filling and
SSA selection included, is re-applied. Realizations are generated with the algorithm of
\citet{emmanoulopoulos2013}, which matches the power spectral density (PSD) and the flux
probability density function; the Gaussian method of \citet{timmer1995} is used only as its
first step, because on its own it overestimates significances
\citep[e.g.][]{otero2023,penil2026}. Each realization is produced on a grid six times longer
than the baseline and four times finer than the bin width, resampled at the observed epochs,
and given Gaussian noise with the observed uncertainties. The search band is 1--6\,yr, as in
\citet{rico2025} and \citet{penil2026}, and global $p$-values are obtained from the maximum of
each simulated periodogram over that band, which requires no count of independent frequencies.

The PSD is fitted with a power law,
\begin{equation}
P(\nu)=A\nu^{-\beta}+C,
\end{equation}
where $A$ sets the red-noise normalization, $\beta$ the slope, and $C$ a frequency-independent
(Poisson-like) noise floor, and with a bending power law,
\begin{equation}
P(\nu)=\frac{A\nu^{-\beta_1}}{1+(\nu/\nu_{\rm b})^{\beta_2-\beta_1}}+C,
\end{equation}
with a break frequency $\nu_{\rm b}$ and a second slope $\beta_2$. Both are fitted by maximizing
the Whittle likelihood \citep{vaughan2010}. They are indistinguishable by the Akaike
information criterion ($\Delta{\rm AIC}=1.46$), and we retain the power law for parsimony, with
$\beta=1.13$, within the range found for gamma-ray blazars by \citet{bhatta2020}. Because this
does not select a null model, we repeat the significance calculation under the bending model and
under $\beta=0.6$ and $\beta=1.8$, each with $2\times10^{4}$ realizations: the global $p$-value
of the candidate moves between 0.14 and 0.15. We do not add autoregressive nulls, because
\citet{penil2026} find ARFIMA calibrations to agree closely with power-law ones and ARIMA to be
the anti-conservative outlier.

These nulls are stationary, whereas the variability of PKS\,0903$-$57 is concentrated in the
second half of the record (Sect.~\ref{sec:data}). We therefore add one non-stationary control.
We write the light curve as $F(t)=E(t)\,s(t)$, where $E$ is the SSA trend (RC0+RC1,
Sect.~\ref{sec:harm}), and generate each realization as an Emmanoulopoulos series matching the
PSD ($\beta=1.06$) and flux distribution of $s=F/E$, multiplied by $E$ and given the observed
measurement noise. From $2\times10^{4}$ realizations the global $p$-value of the SSA statistic
is 0.026 ($2.2\sigma$), that of the raw-flux GLS 0.94, and the joint test of
Sect.~\ref{sec:robust} gives $p_{\rm joint}=0.063\pm0.002$ ($1.9\sigma$). This null makes the
candidate more significant, because a two-year structure confined to the active phase is rarer
in envelope-modulated red noise than in stationary red noise; it does not reach $3\sigma$, and
it does not affect the temporal tests of Sect.~\ref{sec:robust}, which ask whether the structure
persists. Over the nulls tested, the global $p$-value of the SSA statistic ranges from 0.026 to
0.15.

Each realization inherits the observed epochs and gap pattern, and the test-statistic censoring
is therefore treated as fixed, not flux dependent. Because low states are the ones most likely
to fail a test-statistic cut, we test this directly by relaxing the cut to ${\rm TS}\ge1$, which
retains 216 of the 221 bins: the candidate is recovered at 2.12\,yr with the same pair and a
global $p$-value of 0.147 from $2\times10^{4}$ realizations, against 0.146 for the baseline
selection.

For the template analysis \citet{maldonado2026} use Timmer-Koenig realizations of a power law
with a Poisson floor. For the templates applied to the 30-day oscillatory component we use the
Emmanoulopoulos realizations instead, which gives all rows of Table~\ref{tab:results} one
common null and also suits the statistic: the template periodogram measures the fraction of
variance removed by the fit, and a Gaussian red-noise series spreads its variance over the
whole baseline, where a long-period template removes a large fraction of it, whereas a blazar
light curve concentrates it in a few flares. For the blind-search configuration, which
reproduces that work, we adopt its null and quote the Emmanoulopoulos result alongside
(Sect.~\ref{sec:ftp}).

The false-alarm probability is the quantity the simulations measure, and we quote it as the
primary statistic. We report a local value, from the fraction of realizations exceeding the
observed power at the frequency of the peak, and a global value, from the fraction whose maximum
anywhere in the band exceeds it. Because this is a targeted analysis of one source, our global
value includes the search over periods but no trial factor for a sample; it corresponds to the
per-source significance of \citet{maldonado2026}, whose global significance also includes the
number of sources searched. Gaussian equivalents are given in parentheses as two-sided values,
following \citet{rico2025} and \citet{penil2026}, with a global significance below $1\sigma$
quoted as $<1\sigma$; for the blind-search configuration, whose source uses one-sided values, we
quote only false-alarm probabilities. The number of realizations sets a floor $1/(N_{\rm sim}+1)$
on the attainable $p$-value, $5\times10^{-6}$ for the $2.0\times10^{5}$ realizations of the
raw-flux periodogram and $2\times10^{-5}$ for the $5\times10^{4}$ realizations of the SSA-based
statistics; no value reported below approaches it.

\subsection{Harmonic analysis}
\label{sec:harm}

\begin{figure}
\centering
\includegraphics[width=\linewidth]{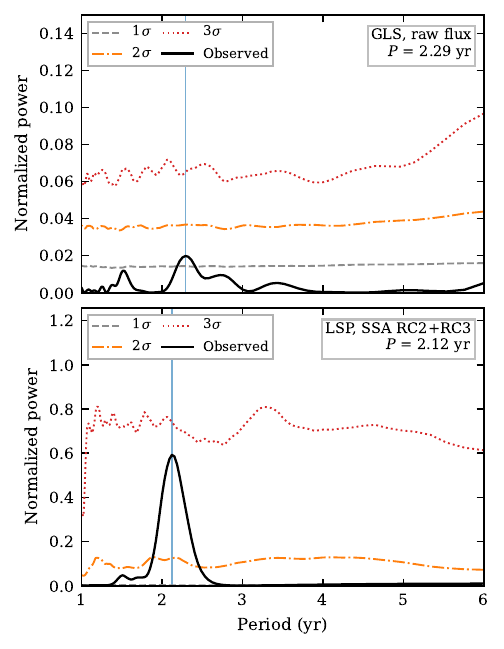}
\caption{Periodograms of the 30-day light curve. Upper panel: generalized Lomb-Scargle
periodogram of the raw flux. Lower panel: Lomb-Scargle periodogram of the SSA oscillatory
component (RC2+RC3). Dashed, dash-dotted and dotted curves are the 1, 2 and $3\sigma$ local confidence
levels, drawn from an independent set of $2\times10^{3}$ Emmanoulopoulos realizations.}
\label{fig:pgram}
\end{figure}

We first compute the generalized Lomb-Scargle periodogram of the raw 30-day flux series
\citep{zechmeister2009}. It peaks at $P=2.29$\,yr with a normalized power of 0.020. From $2.0\times10^{5}$ realizations the
local false-alarm probability is $p=0.19$ ($1.3\sigma$) and the global one $p=0.98$
($<1\sigma$): the raw-flux periodogram is entirely consistent with red noise
(Fig.~\ref{fig:pgram}, upper panel).

We then apply SSA with the mode selection of \citet{dinesh2026}, which is also the one used for
the time-frequency analysis of Sect.~\ref{sec:tfm}; both analyses therefore use the same oscillatory component. Because the selection is data adaptive and is part of the statistic, we
summarize it; the details are given in that work. The 30-day light curve is placed on its full
grid of 221 bins, with the non-detections filled by linear interpolation, embedded with window
length $L=0.4N=88$ \citep{rico2025}, and decomposed by singular value decomposition into
reconstructed components RC0, RC1, \ldots{} Components whose zero-crossing timescale exceeds
6\,yr are assigned to the trend. Each remaining component within the signal range defined by
the tail of the eigenvalue spectrum becomes a candidate if its periodogram peaks inside
1--6\,yr at more than three times its median power and 1.5 times its 90th percentile. Adjacent candidates are accepted as a pair, or as a block of pairs at a
common period, if their W-correlation exceeds a threshold $\rho_{\min}$ set by the
noise-dominated components of the light curve and their eigenvalues and periods are consistent.
The accepted mode with the largest mean eigenvalue per component is the oscillatory component.
Its reconstruction, evaluated at the detected epochs only, is the series analyzed below with a
Lomb-Scargle periodogram over 1--6\,yr on a frequency grid oversampled by a factor of eight.
In the Monte Carlo realizations the mode is re-selected each time, and a realization in which no
mode is accepted contributes zero power.

\begin{figure}
\centering
\includegraphics[width=\linewidth]{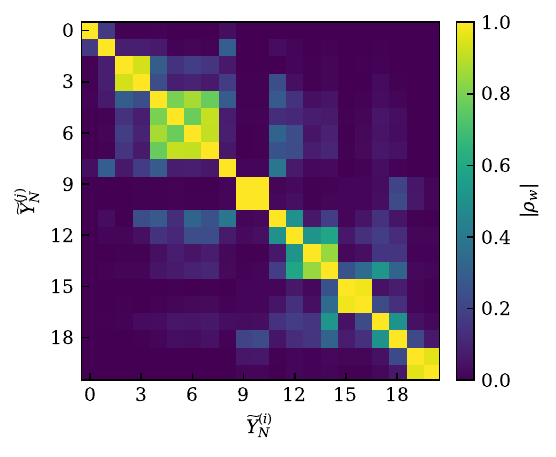}
\caption{Absolute W-correlation matrix $\rho_w(\widetilde Y_N^{(i)},\widetilde Y_N^{(j)})$ of
the reconstructed components RC0--RC20 of the gap-filled 30-day light curve, with $\rho_w=1$
for components that must be grouped together and $\rho_w=0$ for uncorrelated ones. RC0 and RC1
form the trend, and the isolated $2\times2$ block RC2+RC3 is the pair selected as the
oscillatory component by the algorithm of Sect.~\ref{sec:harm}. The $4\times4$ block RC4--RC7
is the next accepted mode, at 1.54\,yr, with a smaller mean eigenvalue per component.}
\label{fig:wcorr}
\end{figure}

For PKS\,0903$-$57 the trend consists of RC0 and RC1, and eight modes are accepted. The selected
mode is the pair RC2+RC3, carrying 6.9 per cent of the eigenvalue sum (Fig.~\ref{fig:wcorr}). The
next is the block RC4--RC7 at 1.54\,yr, which carries more of the eigenvalue sum, 10.7 per cent,
but whose mean eigenvalue per component is 23 per cent smaller. The Lomb-Scargle
periodogram of the RC2+RC3 reconstruction peaks at $P=2.12$\,yr with a normalized power of
0.592; the selection step, on its coarser grid, gives 2.13\,yr. From $5\times10^{4}$ realizations
the global false-alarm probability is $p=0.15$ ($1.5\sigma$), from 7288 exceedances, and the
local one is $p=6.0\times10^{-3}$ ($2.7\sigma$), from 299 exceedances (Table~\ref{tab:results}).
The global value is the quantity on which our conclusion rests. The quantity quoted after each
period throughout this paper is the half width at half maximum of the periodogram peak; it
measures the period resolution of the analysis and is not a statistical uncertainty on the
period. The selected component is dominated by this single peak: the next local maxima, at 1.50
and 1.72\,yr, carry 8 and 6 per cent of its power (Fig.~\ref{fig:pgram}, lower panel).

The candidate appears only after this data-adaptive decomposition: the raw-flux periodogram
shows no corresponding feature (Fig.~\ref{fig:pgram}). We checked whether the decomposition
itself favors peaks near the candidate period, which would bias a test centered on it. In the
realizations the SSA peak period has a median of 1.66\,yr and 10.5 per cent fall within 10 per cent of 2.12\,yr: no such preference is present. This check concerns the location of the
peaks only; the global $p$-value, which already incorporates the full distribution of peak
heights over the band, remains the quantity on which we rely.

\begin{table*}
\caption{Gamma-ray periodicity results, all from the 30-day product.}
\label{tab:results}
\centering
\begin{tabular}{@{}llccccc@{}}
\hline\hline
Estimator & Series & $P$ (yr) & Power & $p_{\rm local}$ & $p_{\rm global}$ & \\
\hline
GLS                    & Raw flux         & 2.29 & 0.020 & $1.9\times10^{-1}$ & $9.8\times10^{-1}$ & ($<1\sigma$) \\
LSP                    & SSA oscillatory  & 2.12 & 0.592 & $6.0\times10^{-3}$ & $1.5\times10^{-1}$ & ($1.5\sigma$) \\
Template, sinusoidal   & SSA oscillatory  & 2.12 & 0.592 & $6.0\times10^{-3}$ & $1.5\times10^{-1}$ & ($1.5\sigma$) \\
Template, Gaussian     & SSA oscillatory  & 2.12 & 0.250 & $9.0\times10^{-3}$ & $2.1\times10^{-1}$ & ($1.3\sigma$) \\
Template, archetype    & SSA oscillatory  & 2.15 & 0.502 & $6.5\times10^{-3}$ & $1.5\times10^{-1}$ & ($1.4\sigma$) \\
WWZ ridge              & Detrended flux   & $1.90^{+0.10}_{-0.13}$ & \ldots & \ldots & \ldots & \\
\hline
\end{tabular}
\tablefoot{The last column repeats the global false-alarm probability as a two-sided
Gaussian equivalent. All estimators use the 1--6\,yr band and Emmanoulopoulos realizations of
the same null: $2.0\times10^{5}$ for the GLS row and $5\times10^{4}$ for the four SSA-based
rows, which share the same realizations; no value is at the floor. The sinusoidal template
reproduces the LSP row by construction, as described in Sect.~\ref{sec:ftp}. The WWZ row is the mean ridge period of the morphological analysis of \citet{dinesh2026}
(Table~\ref{tab:wwz}), measured on the light curve after subtraction of the SSA trend; no
periodogram significance is attached to it, and its interval is the flux-randomization
interval of that analysis. Elsewhere in this paper the quantity following a period is a peak
width, as defined in Sect.~\ref{sec:harm}.}
\end{table*}

\subsection{Template periodogram}
\label{sec:ftp}

Because gamma-ray blazar light curves are dominated by asymmetric flares, a sinusoidal basis
can disperse the power of a genuine non-sinusoidal signal across many frequency bins.
\citet{maldonado2026} applied a Fast Template Periodogram (FTP) to 101 LAT light curves and
showed that the recovered significance can depend strongly on the assumed profile: for
4FGL\,J0102.8+5824 the per-source significance is $0.24\sigma$ with a sinusoid and $3.85\sigma$
with a Gaussian burst. A gain of that size is reached by about one per cent of their noise
realizations, the burst template is the one most affected by isolated flares absent from
their null, and they do not regard that source as a periodicity candidate. We therefore ask whether a morphology-aware estimator recovers more power in
PKS\,0903$-$57 than the sinusoidal one does.

We use the three template families of that work: a sinusoid, a Gaussian burst of fractional
width $\sigma=0.05$, and the empirical archetype obtained by epoch-folding the SSA-detrended
fixed-index 7-day light curve of PG\,1553+113, after the quality cuts of Sect.~\ref{sec:data}, at $P=2.10$\,yr into 15 phase bins and smoothing with a cubic
spline. Each template is represented by a truncated Fourier series with ten harmonics, which
reproduces the Gaussian profile to better than 0.1 per cent in full width at half maximum, and
the phase is scanned on a grid of 128 points. At each trial frequency the amplitude, phase and offset are optimized, which makes the statistic invariant to the template normalization.

Our question is narrower than that of \citet{maldonado2026}, whose blind search retains the
full high-frequency variance of a 7-day series: we ask whether the candidate of
Sect.~\ref{sec:harm} is better described by a burst profile than by a sinusoid, and we therefore
apply the templates to the same 30-day SSA oscillatory component. The sinusoidal template then
reduces to the Lomb-Scargle periodogram of that series, because for a one-harmonic model the two
statistics are the same fraction of variance removed; numerically they agree to
$3\times10^{-4}$ in normalized power at every frequency. Any gain from a burst-shaped template is
therefore measured directly against the harmonic result. The blind-search configuration is
applied separately below.

\begin{figure*}
\centering
\includegraphics[width=0.86\textwidth]{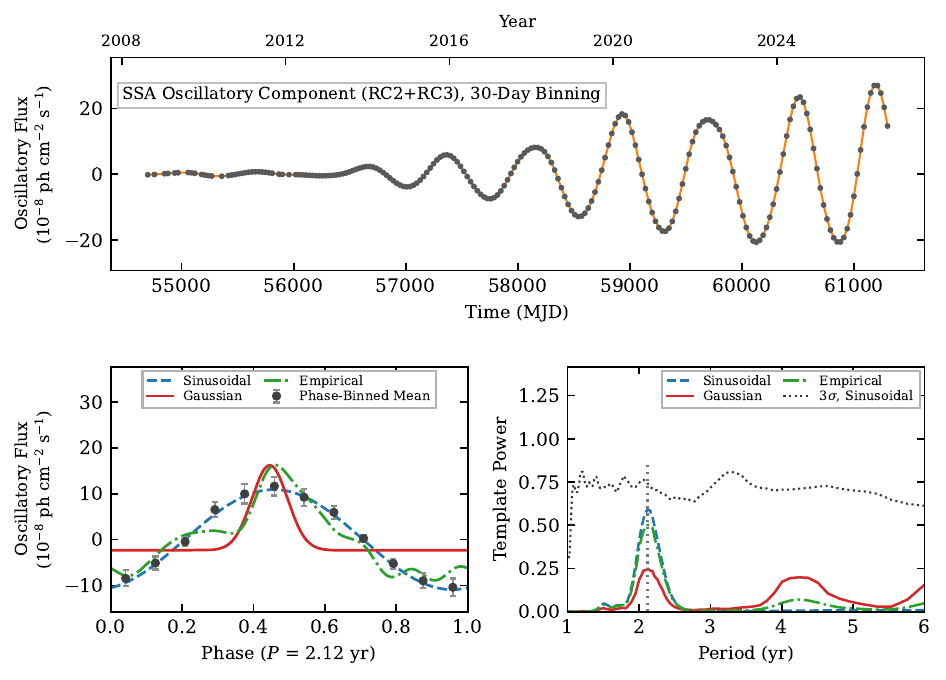}
\caption{Template analysis of the SSA oscillatory component (RC2+RC3) of the 30-day light curve, that is the
flux of Fig.~\ref{fig:lc} after subtraction of the SSA trend. Upper
panel: the oscillatory component, whose amplitude is confined to the second half of the record.
Lower left: the component folded on 2.12\,yr, with phase-binned means and the best-fitting
amplitude, phase and offset of each template at that period; with only a few
cycles of significant amplitude, a coherent-looking folded profile does not by itself indicate a persistent periodicity. Lower right: the three template
periodograms, with the $3\sigma$ local confidence level for the sinusoidal template, drawn
from an independent set of $2\times10^{3}$ Emmanoulopoulos realizations; the dotted vertical line marks 2.12\,yr.}
\label{fig:templates}
\end{figure*}

There is no such gain (Fig.~\ref{fig:templates}, Table~\ref{tab:results}). The sinusoid gives
the highest power of the three, 0.592 at 2.12\,yr, with a global false-alarm probability
$p=0.15$. The empirical archetype gives 0.502 at 2.15\,yr and also $p=0.15$, and the Gaussian
burst gives 0.250 at 2.12\,yr and $p=0.21$. In the notation of \citet{maldonado2026},
$\Delta\sigma=\sigma_{\rm best}-\sigma_{\rm sine}=0.00$: the sinusoid is the best-fitting
template, as it is for 41 of the 101 sources of that study. Because $\Delta\sigma>0.5$ occurs in
37 per cent of their noise realizations, a small positive value would not have indicated a
non-sinusoidal morphology either. No template recovers more power than the sinusoid: spectral
leakage does not hide a burst-dominated signal in this source.

We apply the blind-search configuration of \citet{maldonado2026} as published. It uses the
7-day product of Sect.~\ref{sec:data} and removes only the
secular baseline, namely the leading reconstructed components with zero-crossing timescales
above 6\,yr, estimated on the regular grid with the non-detections filled by interpolation, which retains the high-frequency variance; two components are removed. The logarithmically
binned periodogram of the light curve has a dynamic range of 411 and a fitted index
$\alpha=0.77\pm0.03$; the null is therefore well determined in the sense of that work. The largest
template power falls at 5.13\,yr for the sinusoid and the archetype and at 5.50\,yr for the
Gaussian, at or next to the long-period edge of the 1--5.5\,yr search grid, where that work
anticipates red-noise leakage from the detrending boundary. Near the candidate the three periodograms have at most weak local maxima, at 2.18\,yr, and
the power at 2.12\,yr is below 0.01 in all three.
Against the null of that work, Timmer-Koenig realizations of a power law with a Poisson floor,
the band maxima have single-template false-alarm probabilities of 0.71, 0.34 and 0.76 for the
sinusoid, Gaussian and archetype, and 0.98, 0.71 and 0.99 after the correction for three
templates, the per-source values of that work, from $2\times10^{3}$ realizations as there; none of the observed maxima
exceeds the median of its null distribution once the correction is applied.

The Emmanoulopoulos realizations used elsewhere in this paper behave differently in this
configuration: none of $2\times10^{3}$ reaches the observed band maximum for any template. We do
not read this as evidence for a period at the edge of the grid. The realizations are stationary,
whereas the observed light curve concentrates its variance in the second half of the record
(Sect.~\ref{sec:data}), and once only the components above 6\,yr are removed a long-period
template absorbs that concentration. Consistently, the band-edge maximum is recovered in neither
half of the light curve (see below). Under either null the blind search gives no support to the
2.12\,yr candidate, and the feature it does return is a further expression of the
non-stationarity diagnosed in Sect.~\ref{sec:robust}.

We did not apply the Random Forest classifier of \citet{maldonado2026}, whose score that work
uses as a diagnostic of the sample: it responds to isolated flares as well as to periodic
signals and has no well-defined role in a single-source study. One of its input features, the
split-half period ratio, is defined for a single source: with the same detrending applied to
each half of the 7-day light curve, the best template period is 1.54\,yr in the first half and
1.44\,yr in the second ($r_{\rm period\,splits}=0.93$). Both halves are close to the 1.5\,yr
mode of Sect.~\ref{sec:harm} and far from the candidate; the value depends on the detrending of
each half, and we do not use it further.

\subsection{Time-frequency morphology}
\label{sec:tfm}

Whether a candidate is a persistent periodicity or a transient episode is a question about the
time-frequency plane, not about integrated power. We apply the morphological analysis of
\citet{dinesh2026}, in which PKS\,0903$-$57 is one of the analyzed light curves; the values below
are those of that work, from the same light curve and the same SSA mode as Sect.~\ref{sec:harm}.
SSA supplies the period of the selected mode, $P_{\rm SSA}$, and the trend. The WWZ
\citep{foster1996}, with decay constant $c=0.0125$, is then computed on the detrended light
curve, $F_{\rm det}$, the light curve after subtraction of the trend, and its ridge, the curve
of maximum power in the time-period plane, is traced from $P_{\rm SSA}$, following at each epoch
the local power maximum within $\pm40$ per cent of the starting period. The map is computed for
$F_{\rm det}$ and not for the SSA reconstruction, because the reconstruction would smooth away
the period drift and jitter that the parameters are meant to measure. A ridge point is active
when the WWZ power of $F_{\rm det}$ at the ridge period is at least the local noise floor of
the same map, and the longest contiguous run of active points outside the cone of influence is
retained. The ridge is characterized by a cycle-weighted coherence $Q$, which combines the
strength of the modulation above the noise floor with its longevity; a fractional drift
$\epsilon$; a period volatility $\sigma_{\ln P}$, an operational measure of cycle-to-cycle
stability; the number $N_{\rm active}$ of active cycles; and a persistence $\Pi$. Cycles are
counted only outside the cone of influence, of half-width $\lambda P$ with
$\lambda=1/(2\pi\sqrt{c})\simeq1.42$, and the corrected persistence is
$\Pi=N_{\rm active}/(T_{\rm obs}/P-2\lambda)$. Uncertainties come from 1000 flux-randomization
realizations, and the classification uses the one-sided worst-case bounds
$\Delta X=\max(\Delta X_{\rm lo},\Delta X_{\rm hi})$. The classification has three ranks of
decreasing compatibility with a strictly periodic signal: Rank~I, a stable and coherent modulation; Rank~II, a coherent and
long-lived modulation with an evolving period; and Rank~III, the residual, divided into
subclasses by the first criterion a source fails (Table~\ref{tab:wwz}).

\begin{figure}
\centering
\includegraphics[width=\linewidth]{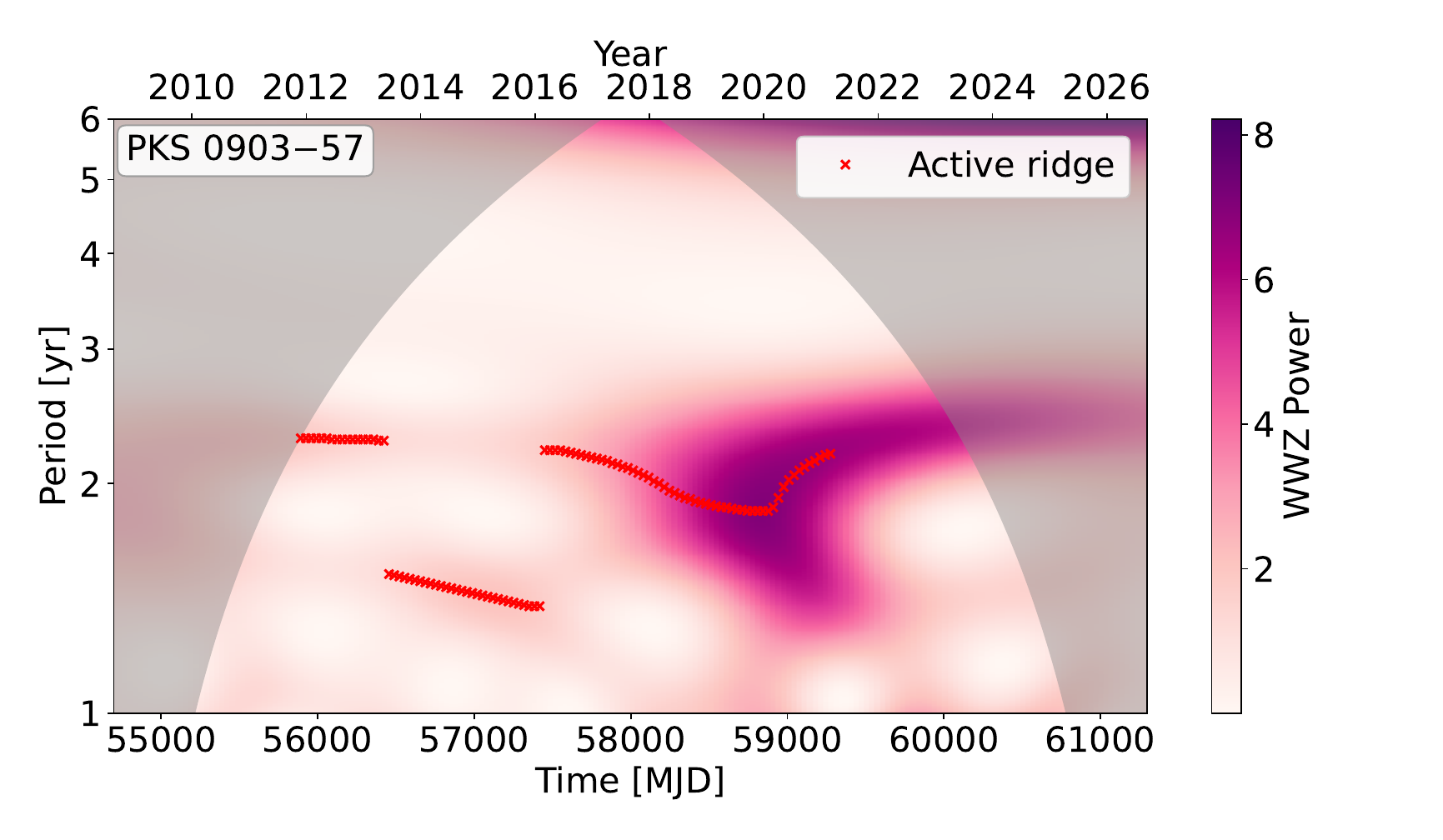}
\caption{WWZ map of the 30-day light curve of PKS\,0903$-$57 after subtraction of the SSA trend,
from \citet{dinesh2026}. The shaded region is the cone of influence and the crosses mark the
active points of the ridge, traced from $P_{\rm SSA}=2.13$\,yr.}
\label{fig:wwz}
\end{figure}

The map is shown in Fig.~\ref{fig:wwz} and the parameters in Table~\ref{tab:wwz}. The ridge is
coherent, with $Q=6.6\pm1.0$ against the gate $Q-\Delta Q>3$, and long-lived, with
$N_{\rm active}=5.1$ cycles and $\Pi=0.76$, and the source passes the uncertainty-quality
selection of that work. Its period, however, is not stable. The mean ridge period is
$1.90^{+0.10}_{-0.13}$\,yr, below $P_{\rm SSA}=2.13$\,yr, and the volatility,
$\sigma_{\ln P}=0.18$, exceeds the boundary $\sigma_{\ln P}+\Delta\sigma_{\ln P}<0.15$ that
\citet{dinesh2026} place between an evolving period and plasma-level volatility. The drift,
$+0.017$\,yr$^{-1}$, is consistent with zero. The source is therefore in Rank~III, subclass
III$_{\sigma}$, which collects coherent candidates with plasma-level volatility; in the sample of
that work the volatility of this subclass is consistent with the rate at which red noise
exceeds the same boundary.

Figure~\ref{fig:wwz} shows where the volatility comes from. The strongest power of the map lies
at 1.8--2.4\,yr between about 2018 and 2022, where the ridge descends from 2.2 to 1.85\,yr and
returns to 2.2\,yr. Before 2016, where the power is low, the ridge follows a weaker feature
near 1.5\,yr, which lies inside its $\pm40$ per cent window, preceded by a short segment near
2.3\,yr around 2012. The switch between the two features produces the 1.90\,yr mean and the
high volatility. The 1.5\,yr feature corresponds to the second SSA mode, RC4--RC7
(Sect.~\ref{sec:harm}), and it is also the period returned by the periodogram when the first
half of the baseline is analyzed alone or the 2018 flare is masked (Sect.~\ref{sec:robust}). The
time-frequency analysis therefore agrees with the periodogram analysis on the dominant period
and on the epoch at which it is strong, and it shows that the light curve does not hold a
single stable period over the record. Because it uses the same light curve and the same SSA
mode, we treat it as corroborating the temporal tests of Sect.~\ref{sec:robust} and not as an
independent result. Because the decomposition contains two modes with comparable eigenvalues,
which of them is selected, and with it the classification, can depend on the baseline
\citep{dinesh2026}, and we quote the classification with that caveat.

\begin{table}
\caption{Ridge morphology of PKS\,0903$-$57 from \citet{dinesh2026}.}
\label{tab:wwz}
\centering
\footnotesize
\setlength{\tabcolsep}{3pt}
\begin{tabular}{@{}lcl@{}}
\hline\hline
Quantity & Value & Worst-case bound and outcome \\
\hline
SSA mode & RC2+RC3 & \ldots \\
$P_{\rm SSA}$ (yr) & 2.13 & \ldots \\
Mean ridge period (yr) & $1.90^{+0.10}_{-0.13}$ & \ldots \\
Coherence $Q$ & $6.6\pm1.0$ & $Q-\Delta Q=5.6>3$: pass \\
Drift $\epsilon$ (yr$^{-1}$) & $+0.017^{+0.012}_{-0.046}$ & $0.063\ge0.02$: not rigid \\
Volatility $\sigma_{\ln P}$ & $0.18^{+0.02}_{-0.07}$ & $0.25\ge0.15$: plasma level \\
Active cycles $N_{\rm active}$ & $5.1^{+0.6}_{-0.9}$ & $4.2>3$: pass \\
Persistence $\Pi$ & $0.76^{+0.03}_{-0.14}$ & $0.62>0.60$ \\
Raw persistence & 0.53 & \ldots \\
Classification & Rank~III$_{\sigma}$ & \ldots \\
\hline
\end{tabular}
\tablefoot{Intervals are the 16th and 84th percentiles of 1000 flux-randomization realizations,
all of which recover the ridge. The third column gives the worst-case bound of each quantity
($X\pm\Delta X$, with $\Delta X$ the larger of the two half-widths) and the criterion it is
compared with. Rank~I requires $Q-\Delta Q>3$, $N_{\rm active}>3$,
$|\epsilon|+\Delta\epsilon<0.02$\,yr$^{-1}$, $\sigma_{\ln P}+\Delta\sigma_{\ln P}<0.06$ and
$\Pi-\Delta\Pi>0.90$; Rank~II requires the same coherence and cycle gates, a failed rigidity
test with $\sigma_{\ln P}+\Delta\sigma_{\ln P}<0.15$, and $\Pi-\Delta\Pi>0.60$
\citep{dinesh2026}. The source meets every Rank~II criterion except the volatility bound.}
\end{table}

\section{Robustness and systematics}
\label{sec:robust}

\subsection{Sensitivity to analysis choices}

We ran 34 variants of the analysis; Table~\ref{tab:robust} lists a representative subset that
includes every temporal variant. They test the
location of the peak and the identity of the selected mode, not the significance. Of the 26
methodological variants, namely SSA window lengths between $0.20N$ and $0.50N$, fixed
W-correlation thresholds between 0.4 and 0.9, the eigenvalue-ratio and screening thresholds
of the selection, its rescue step and its 90th-percentile screen, the search band,
test-statistic thresholds between 2 and 16, and the 7-day product, 24 select RC2+RC3 and return
periods between 2.09 and 2.15\,yr, all within the half width of the baseline peak,
$2.12\pm0.19$\,yr. The other two, $L=0.20N$ and the 1.5--6\,yr band, select
modes at 1.28 and 2.62\,yr because RC2 and RC3 then fail the screening against the 90th
percentile of their periodograms, a level that depends on the window and on the band; without
that screen both return RC2+RC3 at 2.12\,yr. The peak location is therefore robust to the
analysis parametrization. The eight variants that change the temporal content instead scatter
between 1.49 and 4.00\,yr. As the next subsection shows, the existence of a coherent 2.1\,yr
signal is not robust to changes of temporal data selection.

\subsection{Diagnostic tests of temporal stationarity}

Each masked variant treats the affected bins as non-detections: as for every gap, they are
bridged by linear interpolation in the SSA step (Sect.~\ref{sec:harm}) and excluded from the
periodogram, and the complete chain, selection included, is re-applied. The masks remove
MJD\,58200--58400 for the 2018 flare (7 bins), MJD\,58900--59000 for the 2020 flare (3 bins),
and MJD\,58900--59900 for the 2020--2022 active period (33 bins), bracketing the flaring
episodes identified in Sect.~\ref{sec:data}. The split-baseline variants restrict the grid to
the stated interval.

The picture changes when the temporal content is altered. The feature is absent from the
first half of the baseline, where the selected mode peaks at 1.49\,yr, and from the first
twelve years, which give 4.00\,yr, whereas the last twelve years give 2.09\,yr. In the second
half alone, nine years long, the periodogram peaks of RC2 and RC3 are too broad to pass the
90th-percentile screen and a 3.28\,yr mode is selected; without that screen the second half
gives 2.01\,yr with RC2+RC3. Masking the 2018 flare moves the period of RC2+RC3 to 1.50\,yr.
Masking the 2020 flare or the whole 2020--2022 active period, which the interpolation bridges,
leaves a period near 2.2\,yr, but it is then carried by a different pair, RC3+RC4. The recovered feature is
therefore driven by the flaring activity of the second half of the record and depends on
individual flares, and it is absent from the first decade of LAT observations. This is
consistent with the WWZ map of Fig.~\ref{fig:wwz}, in which the 2.1--2.3\,yr feature is strong
only after about 2016 and the ridge follows a 1.5\,yr feature earlier, and with the finding of
\citet{penil2025flares} that a single strong flare can create or destroy an apparent
periodicity.

No variant returns a period near one year, a recognized artifact of gapped
\textit{Fermi}-LAT sampling \citep[e.g.][]{penil2025gaps,maldonado2026} that arises when the
spectral window carries power at that frequency. We computed the spectral window of the
retained epochs at both binnings and find no feature within 10 per cent of one year, the
strongest window features lying between 4 and 5.5\,yr at normalized power of about 0.01, and
the gap fractions of 5.9 and 20.5 per cent are far below the 50 per cent regime in which that
effect becomes important.

\begin{table}
\caption{Selected robustness tests of the SSA candidate.}
\label{tab:robust}
\centering
\footnotesize
\begin{tabular}{@{}lc@{}}
\hline\hline
Variant & $P$ (yr) \\
\hline
Baseline (30-day, fixed index, ${\rm TS}\ge4$) & 2.12 \\
$L=0.20N$ / $0.30N$ / $0.50N$ & 1.28\tablefootmark{a} / 2.15 / 2.12 \\
$\rho_{\min}$ fixed at 0.4 / 0.7 / 0.9 & 2.12 / 2.12 / 2.12 \\
Eigenvalue-ratio threshold 0.1 / 0.3 & 2.12 / 2.12 \\
Screening threshold 2 / 4 & 2.12 / 2.12 \\
No 90th-percentile screen & 2.12 \\
Band 1--5.5 / 1.5--6 / 0.5--6\,yr & 2.12 / 2.62\tablefootmark{a} / 2.12 \\
${\rm TS}$ threshold 2 / 9 / 16 & 2.12 / 2.14 / 2.14 \\
7-day binning & 2.12 \\
\hline
Mask 2018 flare & 1.50 \\
Mask 2020 flare & 2.26\tablefootmark{b} \\
Mask 2020--2022 active period & 2.19\tablefootmark{b} \\
First half of the baseline & 1.49\tablefootmark{c} \\
Second half of the baseline & 3.28\tablefootmark{a} \\
\quad without the 90th-percentile screen & 2.01 \\
First 12\,yr & 4.00\tablefootmark{a} \\
Last 12\,yr & 2.09 \\
\hline
\end{tabular}
\tablefoot{All variants select RC2+RC3 except where noted. This table tests the stability of
the recovered period; it does not re-derive its significance, which for the baseline and its
sensitivity to the PSD model, the test-statistic censoring and a non-stationary null is
quantified in Sect.~\ref{sec:sig}.
\tablefoottext{a}{Selected modes RC8+RC9, RC9+RC10, RC11+RC12 and RC12+RC14, in table
order; in the first three, RC2 and RC3 fail the 90th-percentile screen.}
\tablefoottext{b}{Selected pair RC3+RC4.}
\tablefoottext{c}{Selected pair RC2+RC4.}}
\end{table}

\subsection{Multiplicity of the analysis}

The global significances in Table~\ref{tab:results} account for the search over the whole
period band within one estimator. They do not account for the fact that five periodogram
statistics were computed on the same photons: the GLS of the raw flux, the LSP of the SSA
oscillatory component, and the three templates. Because these are strongly correlated, an analytic correction would require modeling that correlation, and we measure it instead. The
five are not independent pieces of evidence: the sinusoidal template is identical to the LSP,
and four of the five are computed on the same SSA series. For
every realization we compute all five, convert each band maximum to a rank-based global
$p$-value using the realizations themselves, and take the smallest as a joint statistic. The
distribution of that minimum over $5\times10^{4}$ realizations gives
$p_{\rm joint}=0.273\pm0.002$, the uncertainty being binomial, against $p=0.15$ for the single
most significant statistic. For reference this corresponds to $1.1\sigma$ on the convention
used above. The value is well below the \v{S}id\'ak estimate of 0.55 for five independent
tests, as expected because the five statistics are strongly correlated, and unlike that
estimate it is measured. The candidate does not approach a $3\sigma$
detection under any of these treatments. If, as in the population studies of
\citet{penil2026}, a sample trial factor were also applied, the global significance would be
consistent with zero; we do not apply one, because this is a targeted analysis of a single source selected on an optical property, not a blind search.

We record which choices were made after inspecting the data. The data conventions, search
bands, SSA parameters and simulation methods were all fixed in advance from the published
prescriptions cited above. The decision to examine flare-masked and split-baseline variants was
made after inspection. We therefore treat them as diagnostic tests of stationarity and do not
fold them into any significance calculation.

\section{Discussion}
\label{sec:disc}

\subsection{Alternatives to a binary}

Two of the three scenarios considered by \citet{goldoni2024} predict no periodicity at all,
and the gamma-ray null result is what they predict: a recoiling black hole carries a constant
offset and no periodicity, and an infalling clumpy BLR \citep{wang2017} produces irregular
line-profile variability on timescales of days, testable with repeated spectroscopy, but no jet
modulation. The same holds for a single black hole with a non-axisymmetric broad-line region
(Sect.~\ref{sec:blr}).

Within the jet, several mechanisms produce transient year-scale quasi-periodicity without any
binary. Geometric modulation of the Doppler factor in a helical or precessing jet produces episodic maxima, not sinusoidal ones, and can switch on and off as the emitting region
propagates \citep[e.g.][]{camenzind1992}. The geometric masking scenario developed by
\citet{madero2026} and \citet{dominguez2026a}, in which a structured jet with a fast spine and
a slower, softer sheath is differentially boosted as the viewing angle varies, predicts exactly
this behavior: the modulation is visible only during the phase windows in which the geometric
envelope is transparent, and it can be intermittent over a long baseline
\citep[see also][]{dominguez2026b}. A modulation that is strong only in one epoch of an 18.1\,yr record, coherent while
present but with a period that wanders between about 1.5 and 2.3\,yr, is qualitatively
consistent with such a jet-level process, and less easily accommodated by an orbital modulation, which would have to persist throughout the record. We have not fitted a masking model to this source, and this is therefore an interpretation, not an inference.

\subsection{Consequences for the line-shift interpretation}

The gamma-ray data neither support nor require a binary interpretation of the offset broad
H$\alpha$ line, and the absence of a persistent gamma-ray period is not evidence against one: the separations that can host the observed broad line imply orbital
periods of decades to centuries, for which no detectable orbital modulation on the LAT timescale
is expected. The most informative discriminants are optical. Follow-up VLT spectroscopy
obtained between December 2024 and March 2025, reported in preliminary form by
\citet{kasai2026}, recovers the broad symmetric H$\alpha$ component with an offset consistent
within $2\sigma$ with the original measurement, and those authors find no significant change
in its position or shape. This provides an initial temporal baseline but cannot yet
discriminate among the scenarios. A longer and more precise spectroscopic series, ideally
including several broad lines, is needed to test whether the broad-line profile stays stable, as both the binary and the recoiling-hole
scenarios predict \citep[e.g.][]{dotti2009,kelley2021,campanelli2007}, or irregularly variable,
as the infalling-clump scenario predicts \citep{wang2017}. Profile stability alone therefore
does not distinguish a binary from a recoiling hole; only the irregular-variability signature of
the clump scenario is cleanly separated from the other two. Deep imaging to search for a
displaced nucleus is the discriminant most specific to the binary and recoil scenarios
individually.

\subsection{Comparison with previous searches}

PKS\,0903$-$57 does not appear among the candidates of \citet{rico2025}, whose analysis used
12\,yr of data. Our split-baseline test reproduces this directly: restricted to the first
twelve years, the analysis returns 4.00\,yr and not 2.12\,yr. The source is also absent
from the study of \citet{penil2026b} and from the reanalysis of \citet{penil2026},
and it is not among the candidates of \citet{maldonado2026} (Sect.~\ref{sec:data}). Our result
is consistent with all of these. In the
morphological search of \citet{dinesh2026} the source is a Rank~III$_{\sigma}$ candidate, coherent
and long-lived but with plasma-level period volatility, and not a stable periodicity; we have not
modified the criteria of that work to accommodate it.

\section{Summary and conclusions}
\label{sec:concl}

We have tested the binary interpretation of the offset broad H$\alpha$ line of PKS\,0903$-$57
\citep{goldoni2024} kinematically and with the 18\,yr \textit{Fermi}-LAT light curve.

\begin{enumerate}
\item The single-epoch offset alone allows year-scale orbital periods. Requiring the broad-line
region, bound to one hole, to fit inside its Roche lobe sets a floor on the observer-frame
orbital period: 233\,yr for the fiducial mass and a radius of 38\,mpc (185\,yr in the source
frame; 205 to 290\,yr over the statistical mass range), raised further by eccentricity. Over
0.4\,dex in mass and 0.22\,dex in radius the floor spans 60 to 810\,yr, and a 2.12\,yr period
would need a broad-line region 1.1 to 1.3\,dex smaller than the radius-luminosity relation
predicts.
\item The raw-flux periodogram is consistent with red noise ($p=0.98$). The SSA component
RC2+RC3 peaks at 2.12\,yr with a global false-alarm probability $p=0.15$ ($1.5\sigma$; 0.14 to
0.15 over the tested power spectral density models), and the joint test across the five
statistics gives $p=0.273\pm0.002$ ($1.1\sigma$). Under a null that follows the observed
variability envelope these become $p=0.026$ ($2.2\sigma$) and $p=0.063$ ($1.9\sigma$), short of
$3\sigma$.
\item The template periodograms show no morphological preference: the sinusoid, the archetype
and the Gaussian burst give $p=0.15$, 0.15 and 0.21, and since the sinusoid is the most
significant of the three, $\Delta\sigma=\sigma_{\rm best}-\sigma_{\rm sine}=0.00$.
\item Wavelet ridge tracking \citep{dinesh2026} finds a coherent modulation over 5.1 active
cycles with an unstable period: 2.1 to 2.3\,yr after about 2016 and 1.5\,yr before, with a mean
of 1.90\,yr and plasma-level volatility (Rank~III, subclass III$_{\sigma}$).
\item The peak location is robust to analysis choices but not to temporal data selection: the
signal is recovered from the last twelve years but not from the first half or the first twelve
years, and masking the 2018 flare moves it to 1.50\,yr. The 7-day blind search of
\citet{maldonado2026} does not support it under either null model.
\item There is no convincing evidence for a gamma-ray quasi-periodic oscillation in
PKS\,0903$-$57; the 2.1\,yr feature is consistent with a transient, non-stationary structure.
Neither the gamma-ray data nor the orbital dynamics provides positive evidence for a binary
origin of the offset, and neither excludes a wide binary, which would not be expected to
produce a detectable orbital modulation on the LAT timescale.
\end{enumerate}

Continued LAT exposure is the most direct time-domain test: a strictly periodic 2.12\,yr
modulation should persist, and by 2032 the post-2017 segment alone will span seven cycles. A
longer spectroscopic series of the broad-line profile, building on \citet{kasai2026}, remains
the most direct discriminant among the scenarios proposed for the offset.

\section*{Data availability}

The \textit{Fermi}-LAT light curves are public and were retrieved from the Light Curve
Repository \citep{abdollahi2023} on 2026 October 3 (30-day) and October 5 (7-day)
(${\rm TS}_{\min}=1$ products,
P8R3\_SOURCE\_V3). All quantities derived from them are reported in the tables and figures of
this paper.

\section*{Declaration of competing interest}
The authors declare that they have no known competing financial interests or personal
relationships that could have appeared to influence the work reported in this paper.

\section*{Acknowledgements}
AD acknowledges support from MCIN/AEI under grant PID2025-171816NB-I00. A. Dinesh acknowledges
support under grant PID2022-138132NB-C42 (MCIN/AEI/10.13039/501100011033/FEDER, EU), a
UCM-Santander fellowship and UCM FEI-24-0. This work was supported by the European Research
Council, ERC Starting grant MessMapp, S.B. Principal Investigator, under contract no. 949555, and
by the German Science Foundation DFG, research grant ``Relativistic Jets in Active Galaxies''
(FOR 5195, grant No. 443220636). This work has made use of public \textit{Fermi}
data obtained from the High Energy Astrophysics Science Archive Research Center (HEASARC),
provided by NASA Goddard Space Flight Center. We specifically acknowledge the \textit{Fermi}
Science Support Center (FSSC) for the provision of the \textit{Fermi}-LAT Light Curve
Repository.

\bibliographystyle{elsarticle-harv}
\bibliography{refs}

\end{document}